\documentclass[twocolumn,journal]{IEEEtran}
\usepackage[T1]{fontenc}
\usepackage{float}
\usepackage{mathrsfs}
\usepackage{amsmath}
\usepackage{amssymb}
\usepackage{graphicx}
\usepackage[unicode=true,
 bookmarks=true,bookmarksnumbered=true,bookmarksopen=true,bookmarksopenlevel=1,
 breaklinks=false,pdfborder={0 0 0},backref=false,colorlinks=false]
 {hyperref}
\hypersetup{pdftitle={Your Title},
 pdfauthor={Your Name},
 pdfpagelayout=OneColumn, pdfnewwindow=true, pdfstartview=XYZ, plainpages=false}

\makeatletter

\providecommand{\tabularnewline}{\\}
\floatstyle{ruled}
\newfloat{algorithm}{tbp}{loa}
\providecommand{\algorithmname}{Algorithm}
\floatname{algorithm}{\protect\algorithmname}

 \let\oldforeign@language\foreign@language
 \DeclareRobustCommand{\foreign@language}[1]{%
   \lowercase{\oldforeign@language{#1}}}

\usepackage[caption=false,font=footnotesize]{subfig}

\makeatother

\begin{document}

\title{Efficient Sensor Management for Multitarget Tracking in Passive Sensor
Networks via Cauchy-Schwarz Divergence}

\author{Yun~Zhu \thanks{Yun Zhu is with the Key Laboratory of Modern Teaching Technology,
Ministry of Education, and the School of Computer Science. Shaanxi
Normal University, Xi'an City, Shaanxi Province, China.}}

\markboth{}{Your Name \MakeLowercase{\emph{et al.}}: Your Title}
\maketitle
\begin{abstract}
This paper presents an efficient sensor management approach for multi-target
tracking in passive sensor networks. Compared with active sensor networks,
passive sensor networks have larger uncertainty due to the nature
of passive sensing. Multi-target tracking in passive sensor networks
is challenging because the multi-sensor multi-target fusion problem
is difficult and sensor management is necessary to achieve good trade-offs
between tracking accuracy and energy consumption or other costs. To
address this problem, we present an efficient information-theoretic
approach to manage the sensors for better tracking of the unknown
and time-varying number of targets. This is accomplished with two
main technical innovations. The first is a tractable information-based
multi-sensor selection solution via a partially observed Markov decision
process framework. The Cauchy-Schwarz divergence is used as the criterion
to select informative sensors sequentially from the candidates. The
second is a novel dual-stage fusion strategy based on the iterated-corrector
multi-sensor generalized labeled multi-Bernoulli filter. Since the
performance of the iterated-corrector scheme is greatly influenced
by the order of sensor updates, the selected sensors are first ranked
in order of their abilities to detect targets according to the Cauchy-Schwarz
divergence, followed the iterated-corrector update. The computation
costs of ranking the sensors are negligible, since the Cauchy-Schwarz
divergence has been computed in the multi-sensor selection procedure.
Simulation results validate the effectiveness and efficiency of the
proposed approach. \end{abstract}


\IEEEpeerreviewmaketitle{}

\section{Introduction}

\IEEEPARstart{S}{ensor} networks composed of moving robots or static
sensing nodes have attracted attentions in various fields, such as
surveillance, scene analysis, and so forth. Multi-sensor multi-target
tracking, one of the most critical low-level techniques, is difficult
in twofold. On the one hand, multi-sensor fusion for multi-target
tracking is challenging due to data association uncertainty between
measurements and targets. On the other hand, intelligent sensor management
is required to report high quality target-related measurements, with
a view to provide a balance in tracking accuracy and energy consumption
of the sensor network \cite{1,2}. Embedded in highly complex multi-target
systems, sensor management is inherently an optimal nonlinear stochastic
control problem and standard optimal control techniques are not directly
applicable. 

A unified approach to addressing the stochastic nature of the multi-target
sensor management is a Bayesian paradigm in which uncertainty is represented
by multi-target probability density functions in a systematic manner
using finite set statistics (FISST) \cite{3}. The (cardinalized)
probability hypothesis density (PHD) \cite{4,5} and the multi-Bernoulli
\cite{6} filters are popular approaches within the FISST framework.
These filters are crude approximations to the Bayes multi-target filter
and were not explicitly designed to estimate target trajectories.
The most advanced FISST-based algorithm, known as the generalized
labeled multi-Bernoulli (GLMB) filter is an analytic solution to the
Bayes multi-target filter that explicitly produces target tracks \cite{7,8}.
The GLMB filter can be implemented with linear complexity in the number
of measurements using Gibbs sampling \cite{9} and has been demonstrated
to be the most efficient, handling in excess of one million tracks
simultaneously from over a billion measurements \cite{10}. Moreover,
the GLMB approach also admits multi-sensor multi-target solutions
that are linear in the sum of the measurements \cite{11}, as well
as multi-scan solutions that are linear in the number of scans \cite{12}.
What distinguishes the GLMB family from other multi-target filters
is that their approximation errors can be captured analytically \cite{8}.

Within the FISST framework, sensor management is treated as an optimization
problem that yields the best sensor control action. Objective functions
proposed for sensor management were mainly task-based and information
measures. Task-based sensor management includes minimization of the
cardinality variance \cite{13,14}, minimization of statistical mean
of the optimal sub-pattern assignment (OSPA) error \cite{15}, maximization
of the posterior expected number of targets \cite{16,17,18}. One
can design a \textquotedblleft task-driven\textquotedblright{} sensor
management strategy that addresses one of these tasks but it may be
poor in addressing the others. Ad hoc methods that address several
tasks by assigning relative weight to each task can also be used \cite{19,20,21}.
However, this requires one to assign relative values to each task.
Our recent work \cite{22} considers the legacy tracks and measurement-updated
tracks separately, to make full use of information involved in the
multi-target posterior density. To deal with a multitude of performance
criteria in a direct manner simultaneously, some researchers have
proposed to use information theoretic measures as objective or reward
functions. In the context of Bayesian estimation, a principled measure
of information gain is the divergence between the predicted and updated
multi-target densities, and divergences such as Kullback-Leibler \cite{23},
and R${\textstyle \acute{e}}$nyi \cite{24} have been widely used.
However, these divergences cannot be computed analytically. In \cite{25}
and \cite{26}, the authors derived closed form Cauchy-Schwarz divergences
for Poisson and GLMB models, providing additional tools to tackle
complex problems in multi-target systems. A special case of this GLMB
sensor management approach has been applied to drone path-planning
\cite{27}. Multi-objective path-planning for multiple agents with
competing objectives has also been developed in \cite{28}. 

Regardless of the objective function, sensor management is essentially
a global combinatorial optimization problem that is extremely challenging
when the sensor network is large (except for the special case where
only one sensor is selected). To alleviate this computational issue,
a spatial non-maximum suppression heuristics was developed in \cite{29},
which requires tuning the parameter of the suppress gate. A suboptimal
solution was proposed in \cite{30}, but it is not suitable for the
sensor selection problem. A subselection problem for each target was
investigated in \cite{31}, and it was solved by maximizing the information
gain of the PHD filter. In this method, the number of activated sensors
is out of the user\textquoteright s control and hence the system may
not meet the communication and real-time constraints. 

In this paper, we consider the sensor management problem for multi-target
tracking in passive sensor networks. Compared with active sensors,
the use of passive sensors is attractive since they cost less and
are robust in extreme environments. At each time step, receivers in
the passive sensor network send their observations regarding the multi-target
state to a central node called the fusion center, which is responsible
for the final inference. However, since transmit and receive antennas
are placed at different locations, measurements collected by passive
sensor networks are typically subjected to noise corruption, missed
detections, and false alarms. Worse, the detection ability of the
passive sensor network deteriorates severely as the distance from
the receiver increases. To deal with these challenges, we adopt the
iterated-corrector multi-sensor GLMB filter as a centralized fusion
scheme and develop an efficient sensor management solution. The main
contributions are as follows. 

First, a novel information-theoretic multi-sensor selection approach
based on the Cauchy-Schwarz divergence is proposed. Typically, a passive
sensor network is realized with either one receiver and multiple transmitters
or a single transmitter combined with multiple receivers. Due to communication
and real-time constraints, the system may need to select a subset
of sensors to report high quality target-related measurements. In
this paper, we adopt the iterated-corrector multi-sensor GLMB filter
as a centralized fusion scheme and develop an efficient multi-sensor
selection solution under the framework of partially observed Markov
decision process (POMDP). A tractable solution is proposed via sequential
selection of the best from the candidates, based on the Cauchy-Schwarz
divergence between the prior and posterior GLMB densities. This approach
is efficient since it is not necessary to search over all possible
sensor combinations and the Cauchy-Schwarz divergence for GLMBs admits
a closed form expression \cite{26}. 

Second, an improved fusion strategy with a modified update is proposed.
As mentioned above, the iterated-corrector multi-sensor GLMB filter
is used as the fusion scheme, which has simple practical implementation.
However, a well-known drawback of the iterated-corrector update is
that it results in different filtering densities depending on the
order of the sensor updates. If the detection ability of the last
sensor is low, the overall performance of the filter degrades. To
deal with this, we propose a dual-stage iterated-corrector update
method. Firstly, the sensors are ranked according to their abilities
to detect targets based on the Cauchy-Schwarz divergence. The computation
involved in this procedure is negligible, since the Cauchy-Schwarz
divergence has been evaluated in the previous multi-sensor selection.
Secondly, the iterated-corrector update is applied in order of the
ranking. The multi-target density is updated by the sensor with weak
detection ability first, and then the sensor with better detection
performance is used in turn. 

The paper is organized as follows. In Section 2, the necessary background
on GLMB recursion is presented and the multi-static passive radar
network is briefly introduced. In Section 3, the proposed approach
with efficient multi-sensor selection and dual-stage iterated-corrector
update is presented in detailed. In Section 4, numerical studies are
given. Finally, Section 5 concludes the paper.

\section{Problem Formulation}

\subsection{Labeled RFS}

In the FISST framework, the target states take values from a state
space $\mathbb{X}$. The multi-target state space is the space of
all finite subsets of $\mathbb{X}$ and is denoted as $\mathcal{F}(\mathbb{X})$.
A multi-target state at each time is modelled as a random finite set
(RFS), a random variable that take values from $\mathcal{F}(\mathbb{X})$.
To address target trajectories in the FISST framework in a principled
manner, the simplest approach is to incorporate labels into the multi-target
state to identify individual targets \cite{3}. A label $\ell\in\mathbb{L}$
is augmented to the state of each target and the multi-target state
is considered as a finite set on $\mathbb{X}\mathbb{\times L}$. Since
some targets may share the same identity, we require that the RFS
modelling the multi-target state on $\mathbb{X}\mathbb{\times L}$
have distinct labels \cite{3}. This is the essence of labeled RFS,
i.e. marked RFS with discrete and distinct marks \cite{7,8}. By convention,
single-target states are represented by lowercase letters, e.g., $x$,
$\mathbf{x}$, while multi-target states are represented by uppercase
letters, e.g., $X$, $\mathbf{X}$. Symbols for labeled states and
their distributions are bolded to distinguish them from unlabeled
ones, e.g., $\mathbf{x}$, $\mathbf{X}$, etc. 

Suppose that at time $k$, there are $N(k)$ target states $\mathbf{x}_{k,1},\ldots,\mathbf{x}_{k,N(k)}$
each taking values in a labeled state space $\mathbb{X}\mathbb{\times L}$,
and $M(k)$ measurements $\mathrm{\textrm{z}}_{k,1},\ldots,\mathbf{\textrm{z}}_{k,M(k)}$,
each taking values in an observation space $\mathbb{Z}$. The set
of target states and measurements are treated as the multi-target
state and multi-target measurement, respectively

\begin{equation}
\mathbf{X}_{k}=\{\mathbf{x}_{k,1},\ldots,\mathbf{x}_{k,N(k)}\}\in\mathcal{F}(\mathbb{X\times L})
\end{equation}

\begin{equation}
\mathbf{\textrm{Z}}_{k}=\{\mathbf{\textrm{z}}_{k,1},\ldots,\mathbf{\textrm{z}}_{k,M(k)}\}\in\mathcal{F}(\mathbb{Z})\text{.}
\end{equation}

\noindent The multi-target Bayes recursion propagates the multi-target
posterior density $\boldsymbol{\mathbf{\pi}}_{k}(\mathbf{X}_{k}|Z_{1:k})$
in time according to the following update and prediction 

\begin{equation}
\boldsymbol{\mathbf{\pi}}_{k}(\mathbf{X}_{k}|Z_{1:k})=\frac{g_{k}(Z_{k}|\mathbf{X}_{k})\boldsymbol{\mathbf{\pi}}_{k|k-1}(\mathbf{X}_{k}|Z_{1:k-1})}{\int g_{k}(Z_{k}|\mathbf{X})\boldsymbol{\mathbf{\pi}}_{k|k-1}(\mathbf{X}|Z_{1:k-1})\delta\mathbf{X}}\label{eq}
\end{equation}

\begin{equation}
\boldsymbol{\mathbf{\pi}}_{k|k-1}(\mathbf{X}_{k}|Z_{1:k-1})=\int\mathbf{f}_{k|k-1}(\mathbf{X}_{k}|\mathbf{\mathscr{\textrm{X}}})\boldsymbol{\mathbf{\pi}}_{k-1}(\mathbf{X}|Z_{1:k-1})\delta\mathbf{X}\label{eq-1}
\end{equation}

\noindent where $\mathbf{f}_{k|k-1}(\cdot|\mathbf{\cdot})$ is the
multi-target transition density that encapsulates the underlying models
of target motions, births and deaths; $g_{k}(\cdot|\mathbf{\cdot})$
is the multi-target likelihood that encapsulates detection uncertainty,
clutter, data association uncertainty, and the usual observation noise.
Note that the integrals in (\ref{eq}) and (\ref{eq-1}) are not ordinary
integrals, but are set integrals. The set integral for a function
$f:\mathcal{F}(\mathbb{X}\mathbb{\times L})\rightarrow\mathbb{R}$
is given by {[}7{]} 

\begin{equation}
\int\mathbf{f}(\mathbf{X})\delta\mathbf{X}=\sum_{i=0}^{\infty}\frac{1}{i!}\int\mathbf{f}(\{\mathbf{x}_{1},\ldots,\mathbf{x}_{i}\})d(\mathbf{x}_{1},\ldots,\mathbf{x}_{i}).\label{eq-2}
\end{equation}

\noindent The multi-target filtering density captures all information
on the multi-target state, such as the number of targets and their
states, at the current time. 

Throughout the paper, we use the standard inner product notation 

\begin{equation}
\left\langle f,g\right\rangle \triangleq\int f(x)g(x)dx\label{eq-3}
\end{equation}

\noindent and the following multi-object exponential notation 

\begin{equation}
h^{X}\triangleq\prod_{x\in X}h(x).\label{eq-4}
\end{equation}

\noindent The inclusion function and the Kronecker delta function
are given to support arbitrary arguments such as sets, vectors, and
integers, as follows:

\begin{equation}
1_{S}(X)\triangleq\begin{cases}
1, & \mathrm{if}~X\subseteq S\\
0, & \mathrm{otherwise}
\end{cases},~\delta_{S}(X)\triangleq\begin{cases}
1, & \mathrm{if}~X=S\\
0, & \mathrm{otherwise}
\end{cases}.\label{eq-5}
\end{equation}

\subsection{Generalized Labeled Multi-Bernoulli}

An important class of labeled RFS is the GLMB family \cite{7,8}.
Under the standard multi-target model, the GLMB is a conjugate prior
that is also closed under the Chapman-Kolmogorov equation. This means
if we start with a GLMB initial prior, then the multi-target prediction
and posterior densities at any time are also GLMB densities. Let $\mathcal{L}(\mathbb{X}\mathbb{\times L})\rightarrow\mathbb{L}$
be the projection $\mathcal{L}(x,\ell)=\ell$ and $\mathcal{\triangle}(\mathbf{X})\triangleq\delta_{|\mathbf{X}|}(|\mathcal{L}(\mathbf{X})|)$
denote the distinct label indicator. A GLMB is an RFS on $\mathbb{X}\mathbb{\times L}$
distributed according to

\begin{equation}
\mathbf{\boldsymbol{\pi}}(\mathbf{X})=\triangle(\mathbf{X})\sum_{c\in\mathbb{C}}w^{(c)}(\mathcal{L}(\mathbf{X}))[p^{(c)}]^{\mathbf{X}}\label{eq-6}
\end{equation}

\noindent where $\mathbb{C}$ is a discrete index set, $w^{(c)}(L)$
and $p^{(c)}$ satisfy:

\begin{equation}
\sum_{L\subseteq\mathbb{L}}\sum_{c\in\mathbb{C}}w^{(c)}(L)=1\label{eq-7}
\end{equation}

\begin{equation}
\int p^{(c)}(x,\ell)dx=1.\label{eq-8}
\end{equation}

\noindent The GLMB density (\ref{eq-6}) can be interpreted as a mixture
of multi-target exponentials, where each term consists of a weight
$w^{(c)}(\mathcal{L}(\mathbf{X}))$ that depends only on the labels
of $\mathbf{X}$, and a multi-target exponential $[p^{(c)}]\mathbf{^{X}}$
that depends on the entire $\mathbf{X}$.

For implementation, it is easier to use the delta-form of the GLMB,
known as $\delta$-GLMB \cite{8}. In fact, a $\delta$-GLMB is a
GLMB with

\begin{equation}
C=\mathcal{F}(\mathbb{L})\times\varXi
\end{equation}

\begin{equation}
w^{(c)}(L)=w^{(I,\xi)}(L)=w^{(I,\xi)}\delta_{I}(L)
\end{equation}

\begin{equation}
p^{(c)}=p^{(I,\xi)}=p^{(\xi)}
\end{equation}

\noindent where $\varXi$ is a discrete space, i.e., it is distributed
according to

\begin{equation}
\mathbf{\boldsymbol{\pi}}(\mathbf{X})=\triangle(\mathbf{X})\sum_{(I,\xi)\in\mathcal{F}(\mathbb{L_{\textrm{+}}})\mathscr{\mathrm{\times\varXi}}}w_{+}^{(I,\xi)}\times\delta_{I_{+}}(\mathcal{L}(\mathbf{X_{+}}))[p_{+}^{(\xi)}]^{\mathbf{X_{+}}}.\label{eq-9}
\end{equation}

\noindent Clearly, the family of $\delta$-GLMB is also closed under
the Chapman-Kolmogorov prediction and Bayes update. It was shown in
\cite{7} that if the multi-target prior is a $\delta$-GLMB with
the form (\ref{eq-9}), the multi-target prediction is also a $\delta$-GLMB
with the following form

\begin{equation}
\mathbf{\boldsymbol{\pi_{+}}}(\mathbf{X_{+}})=\triangle(\mathbf{X})\sum_{(I_{+},\xi)\in\mathcal{F}(\mathbb{L_{\textrm{+}}})\mathscr{\mathrm{\times\varXi}}}w_{+}^{(I,\xi)}\times\delta_{I_{+}}(\mathcal{L}(\mathbf{X_{+}}))[p_{+}^{(\xi)}]^{\mathbf{X_{+}}}\label{eq-36}
\end{equation}

\noindent where

\begin{equation}
w_{+}^{(I_{+},\xi)}=w_{B}(I_{+}\cap\mathbb{B})w_{S}^{(\xi)}(I_{+}\cap\mathbb{L})\label{eq-10}
\end{equation}

\begin{equation}
p_{+}^{(\xi)}(x,\ell)=1_{\mathbb{L}}(\ell)p_{S}^{(\xi)}(x,\ell)+(1-1_{\mathbb{L}}(\ell))p_{B}(x,\ell)\label{eq-11}
\end{equation}

\begin{equation}
p_{S}^{(\xi)}(x,\ell)=\frac{\left\langle p_{S}(\cdot,\ell)f(x|\cdot,\ell),p^{(\xi)}(\cdot,\ell)\right\rangle }{\eta_{S}^{(\xi)}(\ell)}\label{eq-12}
\end{equation}

\begin{equation}
\eta_{S}^{(\xi)}(\ell)=\int\left\langle p_{S}(\cdot,\ell)f(x|\cdot,\ell),p^{(\xi)}(\cdot,\ell)\right\rangle dx\label{eq-13}
\end{equation}

\begin{equation}
w_{S}^{(\xi)}(L)=[\eta_{S}^{(\xi)}(\ell)]^{L}\sum_{I\subseteq\mathbb{L}}1_{I}(L)[q_{S}^{(\xi)}]^{I-L}w^{(I,\xi)}\label{eq-14}
\end{equation}

\begin{equation}
q_{S}^{(\xi)}(\ell)=\left\langle q_{S}(\cdot,\ell),p^{(\xi)}(\cdot,\ell)\right\rangle .\label{eq-15}
\end{equation}

\noindent For a given label set, $I_{+}$, the weight $w_{+}^{(I_{+},\xi)}$
is a product of the weight $w_{B}(I_{+}\cap\mathbb{B})$ of birth
labels $I_{+}-\mathbb{L}=I_{+}\cap\mathbb{B}$ and the weight $w_{S}^{(\xi)}(I_{+}\cap\mathbb{L})$
of surviving labels $I_{+}\cap\mathbb{L}$. The predicted single-target
density for a given label $p_{+}^{(\xi)}(\cdot,\ell)$ is either the
density $p_{B}^{(\xi)}(\cdot,\ell)$ of a newly born target or the
density $p_{S}^{(\xi)}(\cdot,\ell)$ of a surviving target. 

The multi-target posterior is also a $\delta$-GLMB with the following
form 
\begin{equation}
\begin{array}{c}
\boldsymbol{\mathbf{\pi}}(\mathbf{X}|Z)=\triangle(\mathbf{X})\sum_{(I_{+},\xi)\in\mathcal{F}(\mathbb{L})\mathscr{\mathrm{\times\Xi}}}\sum_{\theta\in\Theta}\\
\times w_{+}^{(I,\xi)}\delta_{I}(\mathcal{L}(\mathbf{X}))[p^{(\xi,\theta)}(\cdot|Z)]^{\mathbf{X}}
\end{array}\label{eq-16}
\end{equation}

\noindent where $\Theta$ is the space of mappings $\theta:\mathbb{L}\rightarrow\{0,1,\ldots,|Z|\}$
such that $\theta(i)=\theta(i^{'})>0$ implies $i=i^{'}$, and
\begin{equation}
\begin{array}{l}
w^{(I,\xi,\theta)}(Z)=\\
\frac{\delta_{\theta^{-1}(\{0:|Z|\})}(I)w^{(I,\xi)}[\eta_{Z}^{(\xi,\theta)}]^{I}}{\sum_{(I,\xi)\in\mathcal{F}(\mathbb{L})\times\Xi}\sum_{\theta\in\Theta}\delta_{\theta^{-1}(\{0:|Z|\})}(I)w^{(I,\xi)}[\eta_{Z}^{(\xi,\theta)}]^{I}}
\end{array}\label{eq-17}
\end{equation}

\begin{equation}
p^{(\xi,\theta)}(x,\ell|Z)=\frac{p^{(\xi)}(x,\ell)\psi_{Z}(x,\ell;\theta)}{\eta_{Z}^{(\xi,\theta)}(\ell)}\label{eq-18}
\end{equation}

\begin{equation}
\eta_{Z}^{(\xi,\theta)}(\ell)=\left\langle p^{(\xi)}(\cdot,\ell),\psi_{Z}(\cdot,\ell;\theta)\right\rangle \label{eq-19}
\end{equation}

\begin{equation}
\begin{array}{c}
\psi_{Z}(x,\ell;\theta)=\delta_{0}(\theta(\ell))q_{D}(x,\ell)+\\
(1-\delta_{0}(\theta(\ell)))\frac{p_{D}(x,\ell)g(z{}_{\theta(\ell)}|x,\ell)}{\kappa(z)}
\end{array}.\label{eq-20}
\end{equation}

It is not tractable to exhaustively compute all components first and
then discard those with small weights in the $\delta$-GLMB recursion.
Hence, truncations via the ranked assignment algorithm and the K-shortest
path algorithm have been proposed to find and keep components with
high weights without having to propagate all the components \cite{8}.
More importantly, the truncation error can expressed in closed form,
and that truncation by keeping the highest weighted components minimizes
the $L_{1}$-approximation error \cite{8}. For this reason we adopt
the GLMB filters in our algorithm.

\subsection{Multi-static Passive Radar Network}

\begin{figure}[htbp]
\noindent \begin{centering}
\includegraphics{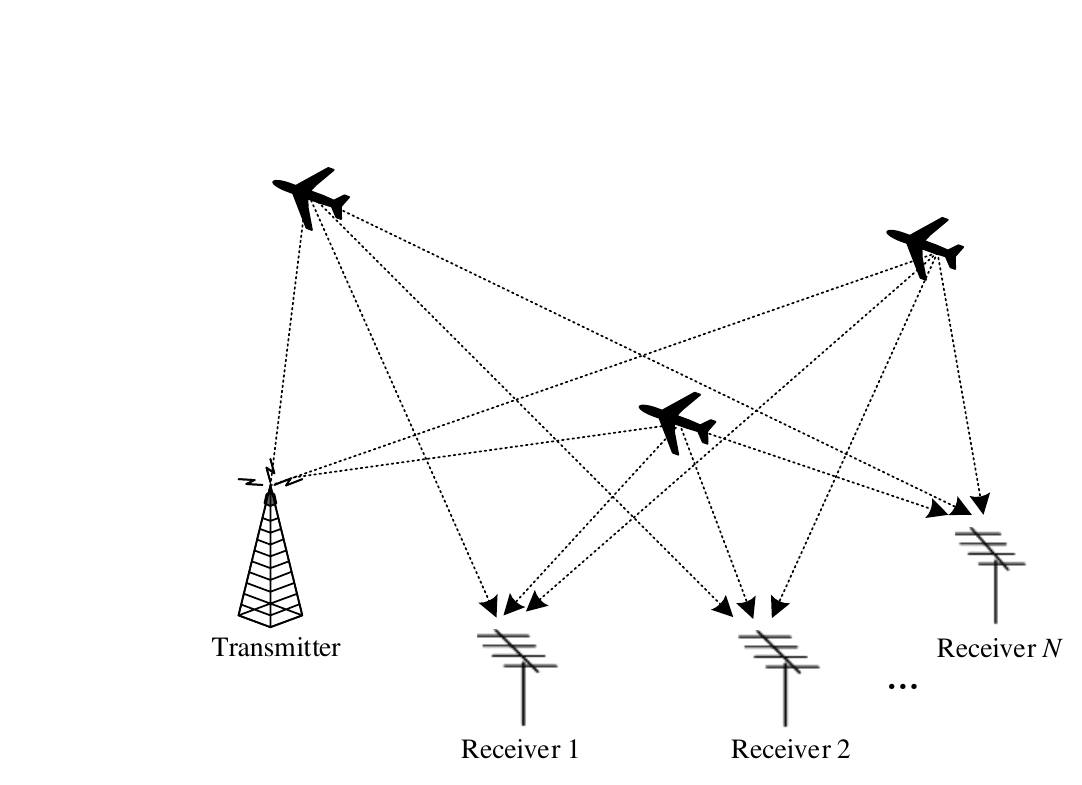}
\par\end{centering}

\caption{Multi-target tracking using a multi-static passive radar system.\label{fig:Multi-target-tracking-using}}
\end{figure}

Tracking of multiple targets in a sensor network is a classical topic
for radar, sonar and other surveillance systems. For target surveillance,
passive radar systems exploit illuminators of opportunity like FM
radio transmitters, digital audio/video broadcasters, WiMAX systems
and global system for mobile-communication (GSM) base stations \cite{32,33,34,35}.
Passive radar systems provide crucial advantages over active systems:
no frequency allocation problem, receivers are hidden for a possible
jamming, energy saving and much lower costs. Typically, a passive
radar sensor network is realized with either one receiver and multiple
transmitters or a single transmitter combined with multiple radar
receivers. The latter usually costs less than the former and is studied
in this paper, as shown in Fig. \ref{fig:Multi-target-tracking-using}. 

In passive radar networks, radar receivers may collect different kinds
of target measurements, such as the time-of-arrival, direction-of-arrival,
bearing, bistatic range, and Doppler shift of the reflected signals.
Target-originated measurements are detected by receiver $j(j=1,2,\ldots,N_{s})$
with probability of detection $p_{D}^{(j)}(x_{k})\leq1$. This probability
is typically a function of the distance between the target in state
$x_{k}$ and receiver $j$ at location $r^{(j)}$. Due to the imperfect
detection of receivers, false detections can also appear. The distribution
of false detections over the measurement space $\mathcal{Z}$ is assumed
time invariant and independent of target states; it will be denoted
by $c^{(j)}(z)$ for receiver $j$. The number of false detections
per scan is assumed to be Poisson distributed, with the constant mean
value $\lambda^{(j)}$. 

At time $k$, the measurement set collected by sensor $j$ is denoted
as $Z_{k}^{(j)}=\{z_{k,1}^{(j)},z_{k,2}^{(j)},\ldots,z_{k,m_{k}^{(j)}}^{(j)}\}$,
where $m_{k}^{(j)}$ is the number of measurements. It is possible
that multiple sensors \textquoteleft simultaneously\textquoteright{}
collect measurements at time $k$. The set of these sensors is referred
to as the set of \textquoteleft activated sensors\textquoteright{}
and is denoted as $A_{k}\subseteq\{1,\ldots,N_{s}\}$. Measurements
from all activated sensors are sent to the fusion centre for processing
as they become available, in the form of messages. A message referring
to measurements collected at time $k$ has the form 

\begin{equation}
\left(k,A_{k},\cup_{j\in A_{k}}\left(j,Z_{k}^{(j)}\right)\right)\label{eq-21}
\end{equation}

\noindent and will be denoted by $Z_{k}^{(A_{k})}$.

\section{Efficient Sensor Management Based on Cauchy-Schwarz Divergence}

\subsection{Sequential Selection of Sensors}

Since the number of sensors in the passive sensor network is generally
large, it would be infeasible to directly use the entire information
of the sensors for tracking multiple targets, and hence sensor selection
is needed. Owing to the geometry, some sensors provide more informative
measurements than others. The decision on which sensors should be
selected can significantly affect the tracking performance. However,
decisions must be made in the presence of uncertainty (both in target
existence and its state) using only the past measurements. Therefore,
the sensor management problem is modeled as a POMDP, as follows

\begin{equation}
\Psi=\{\mathbf{X}_{k},\mathbb{S},\mathbf{f}_{k|k-1}(\mathbf{X}_{k}|\mathbf{X}_{k-1}),g_{k}(Z_{k}|\mathbf{X}_{k}),\vartheta(\mathbf{X}_{k-1},A_{k},\mathbf{X}_{k})\}\label{eq-22}
\end{equation}

\noindent where $\mathbb{S}$ denotes a finite set of sensors for
selection and $\vartheta(\mathbf{X}_{k-1},A_{k},\mathbf{X}_{k})$
is an objective function that measures a reward or cost for transition
from $\mathbf{X}_{k-1}$ to $\mathbf{X}_{k}$ given that the set of
sensors $A_{k}\subseteq\mathbb{S}$ are selected. Note that the general
formulation of a POMDP involves a p-step future decision process,
whereas a single step ahead (myopic) policy is considered in this
paper.

By adopting the information theoretical approach to sensor selection,
the reward function is a measure of \textquoteleft information gain\textquoteright{}
associated with each action. An optimal one-step ahead sensor selection
is formulated as:

\begin{equation}
A_{k}^{*}=\underset{A_{k}\subseteq\mathbb{S}}{\textrm{argmax}}\left\{ \mathbb{E}_{Z_{k}^{(A_{k})}}\left[\vartheta(\mathbf{X}_{k-1},A_{k},\mathbf{X}_{k})\right]\right\} .\label{eq-23}
\end{equation}

\noindent The fact that the reward function $\vartheta$ depends on
the future measurement set $Z_{k}^{(A_{k})}$ is undesirable, since
we want to decide on the future action without actually applying them
before the decision is made. Theoretically, all possible measurement
sets should be used to implement the expectation operator $\mathbb{E}$
in (\ref{eq-23}). To reduce the computational burden, the predicted
ideal measurement set (PIMS) approach \cite{16} is used. First, the
estimated number and states of targets are obtained from the predicted
multi-target density. Then, a measurement is generated for each target
under ideal conditions of no clutter, no measurement noise and perfect
detection. Nevertheless, the computation of the reward function $\vartheta$
can still be expensive. For the case of GLMBs, common information
divergence measures such as the Kullback-Leibler or R$\acute{e}$nyi
divergences \cite{23,24} cannot be evaluated in closed form and Monte
Carlo (MC) integration is often required.

In this paper, the Cauchy-Schwarz divergence is used as the information
measure, since it admits closed form expressions for Poissons \cite{25}
and GLMBs \cite{26}. The Cauchy-Schwarz divergence between the GLMB
prior density and posterior density is used as the reward function,
i.e.

\begin{equation}
\vartheta(\mathbf{X}_{k-1},A_{k},\mathbf{X}_{k})=D(\mathbf{\boldsymbol{\pi}}_{k}(\mathbf{X}|Z_{1:k-1}),\boldsymbol{\pi}_{k}(\mathbf{X}|Z_{1:k-1},Z_{k}^{(A_{k})}))\label{eq-25}
\end{equation}

\noindent where the prior density $\mathbf{\boldsymbol{\pi}}_{k}(\mathbf{X}|Z_{1:k-1})$
is obtained by the GLMB prediction (\ref{eq-36}), and the posterior
density $\boldsymbol{\pi}_{k}(\mathbf{X}|Z_{1:k-1},Z_{k}^{(A_{k})})$
is obtained using the PIMS from the sensors $A_{k}$. If we denote
$\mathbf{\boldsymbol{\pi}}_{k}(\mathbf{X}|Z_{1:k-1})$ and $\boldsymbol{\pi}_{k}(\mathbf{X}|Z_{1:k-1},Z_{k}^{(A_{k})})$
as

\begin{equation}
\boldsymbol{\mathbf{\pi}}_{k}(\mathbf{X}|Z_{1:k-1})=\triangle(\mathbf{X})\sum_{c\in\mathbb{C}}w_{\phi}^{(c)}(\mathcal{L}(\mathbf{X}))[p_{\phi}^{(c)}(\cdot)]^{\mathbf{X}}\label{eq-26}
\end{equation}

\begin{equation}
\boldsymbol{\mathbf{\pi}}_{k}(\mathbf{X}|Z_{1:k-1},Z_{k}^{(A_{k})})=\triangle(\mathbf{X})\sum_{d\in\mathbb{D}}w_{\psi}^{(d)}(\mathcal{L}(\mathbf{X}))[p_{\psi}^{(d)}(\cdot)]^{\mathbf{X}}\label{eq-27}
\end{equation}

\noindent respectively and assume that $p_{\phi}^{c}$ and $p_{\psi}^{d}$
are measured in units of $K^{-1}$, then the Cauchy-Schwarz divergence
between $\boldsymbol{\mathbf{\pi}}_{k}(\mathbf{X}|Z_{1:k-1})$ and
$\boldsymbol{\mathbf{\pi}}_{k}(\mathbf{X}|Z_{1:k-1},Z_{k}^{(A_{k})})$
is
\begin{equation}
\begin{array}{l}
D_{CS}(\boldsymbol{\mathbf{\pi}}_{k}(\mathbf{X}|Z_{1:k-1}),\boldsymbol{\mathbf{\pi}}_{k}(\mathbf{X}|Z_{1:k-1},Z_{k}^{(A_{k})}))=\\
-\ln\left(\frac{\begin{split}\left\langle \boldsymbol{\mathbf{\pi}}_{k}(\mathbf{X}|Z_{1:k-1}),\boldsymbol{\mathbf{\pi}}_{k}(\mathbf{X}|Z_{1:k-1,}Z_{k}^{(A_{k})})\right\rangle _{K}\end{split}
}{\begin{split}\sqrt{\left\langle \boldsymbol{\mathbf{\pi}}_{k}(\mathbf{X}|Z_{1:k-1}),\boldsymbol{\mathbf{\pi}}_{k}(\mathbf{X}|Z_{1:k-1})\right\rangle _{K}}\times\\
\sqrt{\left\langle \boldsymbol{\mathbf{\pi}}_{k}(\mathbf{X}|Z_{1:k-1,}Z_{k}^{(A_{k})}),\boldsymbol{\mathbf{\pi}}_{k}(\mathbf{X}|Z_{1:k-1,}Z_{k}^{(A_{k})})\right\rangle _{K}}
\end{split}
}\right)
\end{array}\label{eq-28}
\end{equation}

\noindent where 

\begin{equation}
\begin{array}{l}
\left\langle \boldsymbol{\mathbf{\pi}}_{k}(\mathbf{X}|Z_{1:k-1}),\boldsymbol{\mathbf{\pi}}_{k}(\mathbf{X}|Z_{1:k-1,}Z_{k}^{(A_{k})})\right\rangle _{K}=\\
\sum_{L\in\mathbb{L}}\sum_{\tfrac{c\in\mathbb{C}}{d\in\mathbb{D}}}w{}_{\phi}^{(c)}(L)w{}_{\psi}^{(d)}(L)\prod_{\ell\in L}K\left\langle p_{\phi}^{(c)}(\cdot,\ell),p_{\psi}^{(d)}(\cdot,\ell)\right\rangle 
\end{array}\label{eq-29}
\end{equation}

\begin{equation}
\begin{array}{l}
\left\langle \boldsymbol{\mathbf{\pi}}_{k}(\mathbf{X}|Z_{1:k-1}),\boldsymbol{\mathbf{\pi}}_{k}(\mathbf{X}|Z_{1:k-1})\right\rangle _{K}=\\
\sum_{L\in\mathbb{L}}\sum_{\tfrac{c\in\mathbb{C}}{d\in\mathbb{C}}}w{}_{\phi}^{(c)}(L)w{}_{\phi}^{(d)}(L)\prod_{\ell\in L}K\left\langle p_{\phi}^{(c)}(\cdot,\ell),p_{\phi}^{(d)}(\cdot,\ell)\right\rangle 
\end{array}\label{eq-30}
\end{equation}

\begin{equation}
\begin{array}{l}
\left\langle \boldsymbol{\mathbf{\pi}}_{k}(\mathbf{X}|Z_{1:k-1,}Z_{k}^{(A_{k})}),\boldsymbol{\mathbf{\pi}}_{k}(\mathbf{X}|Z_{1:k-1,}Z_{k}^{(A_{k})})\right\rangle _{K}=\\
\sum_{L\in\mathbb{L}}\sum_{\tfrac{c\in\mathbb{D}}{d\in\mathbb{D}}}w{}_{\psi}^{(c)}(L)w{}_{\psi}^{(d)}(L)\prod_{\ell\in L}K\left\langle p_{\psi}^{(c)}(\cdot,\ell),p_{\psi}^{(d)}(\cdot,\ell)\right\rangle 
\end{array}.\label{eq-31}
\end{equation}

\noindent Note that $D_{CS}(\cdot,\cdot)$ is invariant to the unit
of hyper-volume $K$. 

Even though the reward function admits a closed form expression, finding
its global optima is still intractable because searching over all
possible sensor combinations is an NP-hard combinatorial optimization
problem \cite{36,37}. We propose a suboptimal multi-sensor selection
method which is computationally tractable and simple to implement.
It is assumed that a fixed number $P$ of sensors are selected at
each time step, that is, $P=|A_{k}|=\textrm{const}$. At time $k$,
$P$ sensors are selected sequentially from the candidates, as follows

\begin{equation}
A_{k}^{(j)*}=\underset{A_{k}^{(j)}\in\mathbb{S}}{\textrm{argmax}}\{\vartheta(\mathbf{X}_{k-1},A_{k}^{(j)},\mathbf{X}_{k})\}\label{eq-32}
\end{equation}

\noindent and

\begin{equation}
A_{k}^{*}=\bigcup_{j=1}^{P}A_{k}^{(j)*}\label{eq-33}
\end{equation}

\noindent where $A_{k}^{(j)*}$ denotes the $j$th selected sensor
and $\mathbb{S}$ denotes the set of remaining candidate sensors which
have not been selected. The computational complexity $\mathcal{O}(N_{s})$
of the proposed approach is much smaller than searching all possible
$P$ sensors combinations in the sensor network which has complexity
$\mathcal{O}(N_{s}!/(N_{s}-P)!)$. These two methods have the same
computation burden and output if and only if $P=1$.

\subsection{Dual-stage Sensor Fusion}

\begin{figure}[htbp]
\noindent \begin{centering}
\includegraphics{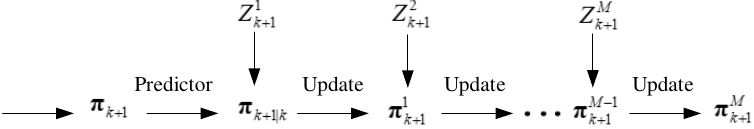}
\par\end{centering}

\caption{Iterated-corrector multi-sensor GLMB filter with one predictor and
multiple correctors.\label{fig:Iterated-corrector-multi-sensor-}}
\end{figure}

For the selected sensors, the iterated-corrector multi-sensor GLMB
update is used in a centralized fusion scheme, as shown in Fig. \ref{fig:Iterated-corrector-multi-sensor-}.
The iterated-corrector multi-sensor GLMB filter consists of one predictor
and multiple correctors. Compared to the single sensor case, the iterated-corrector
multi-sensor filter can reduce the effect of miss detection and false
alarm. It is an attractive multi-sensor algorithm due to its simple
implementation and low computational complexity. However, the iterated-corrector
update invariably yields different tracking results with different
order of sensor updates. For example, a tracking system consists of
$P-1$ \textquoteleft normal\textquoteright{} sensors and a \textquoteleft bad\textquoteright{}
sensor with low detection ability. If the \textquoteleft bad\textquoteright{}
sensor is the $i$th sensor, its negative impacts may be reduced by
the later $P-i$ sensors in the sequential update. If the \textquoteleft bad\textquoteright{}
sensor is the last one to update, the tracking results would be poor,
since there is no sensor to correct the negative impacts caused by
the \textquoteleft bad\textquoteright{} one. In other words, the \textquoteleft bad\textquoteright{}
sensor with a larger value of $i$ would have more negative impact
on the tracking results. 

Compared with active sensor networks, passive sensor networks have
higher uncertainty. The measurements collected by passive sensor networks
are typically subjected to noise corruption, missed detection, and
false alarms. What\textquoteright s more, the probability of detection
of the passive sensor network deteriorates rapidly as the distance
from the receiver increases, making multi-sensor fusion even more
challenging. To address this problem, a novel dual-stage approach
is developed to improve the fusion accuracy without introducing any
additional computation. In the first stage, the selected sensors are
ranked in increasing order of detection abilities. In the second stage,
the iterated-corrector multi-sensor GLMB filter is applied according
to the obtained ranking of selected sensors. The reward functions
obtained in Section 3.1 are used as the ranking criterion since the
value of the reward function can directly reflect the detection ability
of the sensor. The sensor with minimum reward function value is updated
first and the one with the maximum value is updated last. As the update
progresses, the sensor with better detection ability will correct
the previous result. At time step $k$, the pseudocode of the dual-stage
fusion process is presented in Algorithm 1.

\begin{algorithm}[tbh]
\caption{Pseudocode for a single run of the dual-stage multi-sensor GLMB fusion}

$\mathbf{Input:}$

$\boldsymbol{\rightarrow}$selected sensors: $A_{k}^{*}=\cup_{j=1}^{P}A_{k}^{(j)^{*}}$

$\boldsymbol{\rightarrow}$reward function $\vartheta(\mathbf{X}_{k-1},A_{k}^{(j)},\mathbf{X}_{k})$
corresponding to sensor $A_{k}^{(j)*}(j=1,\ldots,P)$ 

$\boldsymbol{\rightarrow}$current multi-Bernoulli density $\mathbf{\boldsymbol{\pi}}_{k-1}(\mathbf{X})=\triangle(\mathbf{X})\sum_{(I,\xi)\in\mathcal{F}(\mathbb{L})\mathscr{\mathrm{\times\varXi}}}w^{(I,\xi)}(\mathcal{L}(\mathbf{X}))[p^{(\xi)}]^{\mathbf{X}}$ 

$\boldsymbol{\rightarrow}$birth model $\mathbf{\boldsymbol{\pi}_{\mathrm{\mathit{B,k}}}}(\mathbf{X}_{\mathrm{\mathit{+}}})=\triangle(\mathbf{X_{+}})\sum_{L\in\mathcal{F}(\mathbb{B})\mathscr{\mathrm{\times\varXi}}}w_{B,k}(L)\times\delta_{L}(\mathcal{L}(\mathbf{X_{+}}))[p_{B,k}]^{\mathbf{X_{+}}}$

$\boldsymbol{\rightarrow}$survival probability function $p_{S}(x,\ell)$

$\boldsymbol{\rightarrow}$single-target transition density $f(x|\cdot,\ell)$
and likelihood $g(z|x,\ell)$

$\boldsymbol{\rightarrow}$sensor model parameters: sensor positions
$s_{j}=[s_{x},s_{y}]^{\textrm{T}}(j=1,\ldots,P)$, detection probability
$p_{D,j}(\cdot)$, and clutter intensity $\kappa(\cdot)$ 

$\mathbf{Output:}$

$\boldsymbol{\rightarrow}$ the fused posterior density to be propagated
to next time, parameterized by $\{\widetilde{r}^{(\ell)},\{\widetilde{w}_{j}^{(\ell)},\widetilde{x}_{j}^{(\ell)}\}_{j=1}^{\tilde{J}^{(\ell)}}\}_{\ell\in\mathbb{L}}$

$\mathbf{Step\ 1:}$ Rank the selected sensors $A_{k}^{*}$ in order
of the value of the reward function $\vartheta(\mathbf{X}_{k-1},A_{k}^{(j)},\mathbf{X}_{k})(j=1,\ldots,P)$
from small to large

$\mathbf{Step\ 2:}$ Iterated-corrector multi-sensor GLMB fusion

\begin{enumerate}
\item $\mathbf{for}$ sensor $j\in A_{k}^{*}$ $\mathbf{do}$
\item $\mathbf{if}$ $j=1$ $\mathbf{do}$
\item Predict the prior $\delta-$GLMB using (\ref{eq-36})-(\ref{eq-15})
to obtain the multi-target prediction $\mathbf{\boldsymbol{\pi}}{}_{+}^{(j)}=\triangle(\mathbf{X{}_{+}})\sum_{(I_{+},\xi)\in\mathcal{F}(\mathbb{L}_{+})\mathscr{\mathrm{\times\varXi}}}w_{+}^{(I,\xi)}\times\delta_{I_{+}}(\mathcal{L}(\mathbf{X_{+}}))[p_{+}^{(\xi)}]^{\mathbf{X_{+}}}$
\item $\mathbf{else}$ $\mathbf{do}$
\item Pseudo-predict $\mathbf{\boldsymbol{\pi}}{}_{+}^{(j)}=\mathbf{\boldsymbol{\pi}}{}^{(j-1)}(\mathbf{X}|Z)$
\item $\mathbf{end\ }\mathbf{if}$
\item Update the predicted $\delta-$GLMB using (\ref{eq-16})-(\ref{eq-20})
to obtain the multi-target posterior $\mathbf{\boldsymbol{\pi}}{}^{(j)}(\mathbf{X}|Z)=\triangle(\mathbf{X})\sum_{(I,\xi)\in\mathcal{F}(\mathbb{L})\mathscr{\mathrm{\times\varXi}}}\sum_{\theta\in\Theta}w^{(I,\xi)}\times\delta_{I}(\mathcal{L}(\mathbf{X}))[p^{(\xi,\theta)}(\cdot|Z)]^{\mathbf{X}}$
\item $\mathbf{end\ }\mathbf{for}$
\item The fused posterior $\mathbf{\boldsymbol{\pi}}(\mathbf{X}|Z)=\mathbf{\boldsymbol{\pi}}{}^{(j)}(\mathbf{X}|Z)$\end{enumerate}
\end{algorithm}

\subsection{Implementation}

Each iteration of the GLMB filter involves an update operation and
a prediction operation, both of which result in weighted sums of multi-target
exponentials with intractably large number of terms. To truncate these
sums and enable an efficient implementation, the ranked assignment
and K-shortest path algorithms are used in the update and prediction,
respectively \cite{8}. In addition, inexpensive look-ahead strategies
can be adopted to reduce the number of computations \cite{8}. There
are two implementations of the GLMB recursion: one is using Gaussian
mixtures (GMs) and the other is using sequential Monte Carlo (SMC)
method. For a linear Gaussian multi-target model (with constant survival
and detection probabilities), each relevant single target density
$p_{k-1}^{(\xi)}(\cdot,\ell)$ is represented as a GM, and the predicted
and updated densities $p_{k|k-1}^{(\xi)}(\cdot,\ell)$, $p_{k}^{(\xi)}(\cdot,\ell)$
are computed using the standard GM update and prediction formulas
based on the Kalman filter. For non-linear non-Gaussian multi-target
models (with state dependent survival and detection probabilities),
each single target density $p_{k-1}^{(\xi)}(\cdot,\ell)$ is represented
by a set of weighted particles. The corresponding predicted and updated
densities $p_{k|k-1}^{(\xi)}(\cdot,\ell)$, $p_{k}^{(\xi,\theta)}(\cdot,\ell)$
are computed by the SMC method. In this paper, the SMC implementation
is adopted to handle the nonlinear dynamic and measurement models,
and also the state-dependent probability of detection.

\begin{figure}[htbp]
\noindent \begin{centering}
\includegraphics{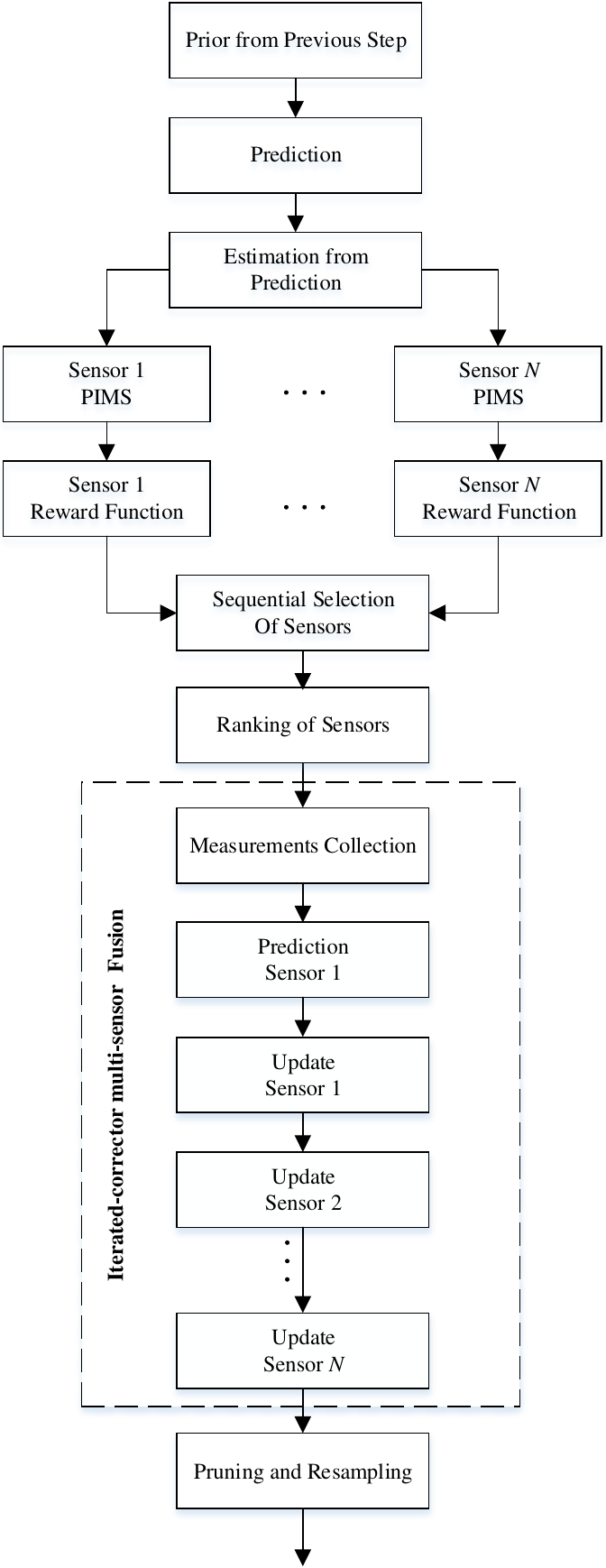}
\par\end{centering}

\caption{Schematic diagram of the ESS-DFS-GLMB filter. \label{fig:Schematic-diagram-of}}
\end{figure}

Starting with a GLMB prior (which is the fused GLMB posterior from
previous time), GLMB filtering is implemented in conjunction with
an efficient sensor selection (ESS) and dual-stage fusion (DSF). A
general schematic diagram for a complete run of the proposed ESS-DFS-GLMB
filter is shown in Fig. \ref{fig:Schematic-diagram-of}.

\section{Simulations}

\begin{figure}[htbp]
\noindent \begin{centering}
\includegraphics{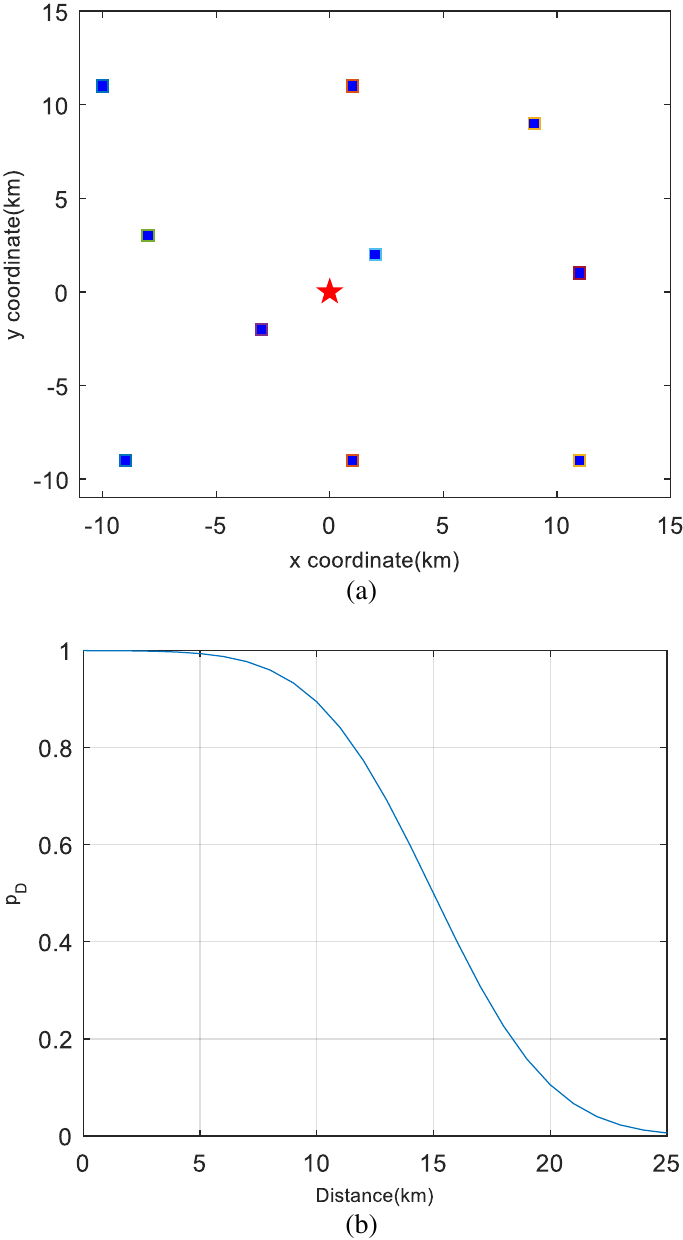}
\par\end{centering}

\caption{System setup: (a) The locations of receivers (squares) and the transmitter
(star); (b) Probability of detection as a function of the distance
from a receiver. \label{fig:System-setup:-(a)} }
\end{figure}

Two challenging scenarios are used to study the performance of the
proposed ESS-DFS-GLMB filter in a passive sensor network. Bistatic
range (transmitter-target-receiver) and bearing measurements are used.
In Scenario 1, a total number of three targets with non-linear motions
move through the surveillance area and in close proximity to each
other. In Scenario 2, a more challenging multi-target tracking problem
is considered. The total number of targets is increased into six,
and locations of targets are relatively scattered. The scattered distribution
of the targets makes it impossible to intuitively obtain the optimal
selection solution and will increase the difficulty of sensor management.
The structure of the passive sensor network is borrowed from \cite{38},
where one transmitter and ten receivers are placed in the $x\text{\textendash}y$
plane as shown in Fig. \ref{fig:System-setup:-(a)}(a). During the
target tracking process, $P$ receivers are selected automatically
at each time step. The probability of detection is modeled as follows
\cite{38}

\begin{equation}
p_{D}^{(j)}(x_{k})=1-\phi(\left\Vert p_{k}-r^{j}\right\Vert ;\alpha,\beta)
\end{equation}

\noindent where $p_{k}=[x_{k},y_{k}]^{\textrm{T}}$ is the target
position, $r^{j}=[x_{R}^{j},y_{R}^{j}]^{\textrm{T}}$ is the position
of receiver $j$, $d_{k,j}=\left\Vert p_{k}-r^{j}\right\Vert $ is
the distance between the target and the receiver, and $\phi(d;\alpha,\beta)=\int_{-\infty}^{d}\mathcal{N}(v;\alpha,\beta)dv$
is the Gaussian cumulative distribution function with $\alpha=15$
km and $\beta=(4\textrm{ km})^{2}$. Fig. \ref{fig:System-setup:-(a)}(b)
plots the probability of detection as a function of the distance between
the receiver and the target. As the distance from the receiver increases,
the probability of detection of the passive sensor decreases quickly
\cite{39}. The poor detection ability makes the sensor management
problem for target tracking difficult, and many state-of-the-art techniques
would fail. 

\begin{figure}[htbp]
\noindent \begin{centering}
\includegraphics{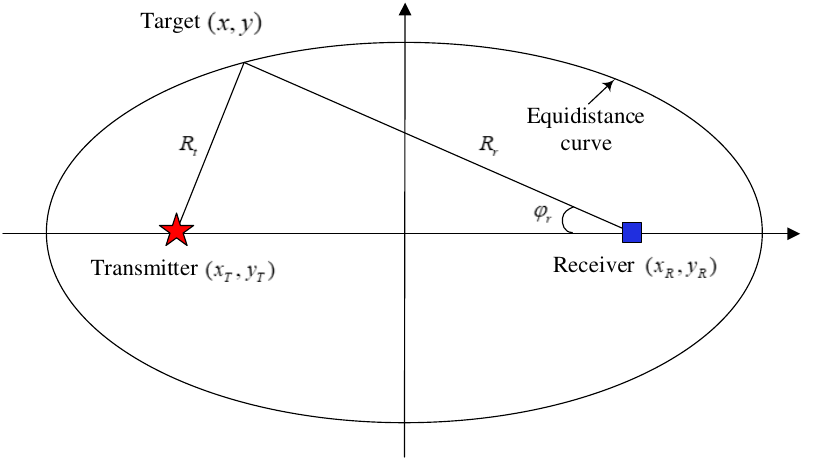}
\par\end{centering}

\caption{The geometry of a single transmitter-target-receiver pair. The equidistance
curve can be plotted using the bistatic range $\rho=R_{t}+R_{r}$
and the target position is further determined using the bearing $\varphi_{r}$.
\label{fig:The-geometry-of}}
\end{figure}

If a target located at $p_{k}=[x_{k},y_{k}]^{\textrm{T}}$ is detected
by receiver $j$, then the target-originated measurement is a noisy
bearing and bistatic range (transmitter-target-receiver) vector given
by 

\begin{equation}
\begin{array}{l}
z_{k}^{j}=\left[\begin{array}{c}
\varphi^{j}\\
\rho^{j}
\end{array}\right]=\\
\left[\begin{array}{c}
\arctan\left(\frac{y_{k}-y_{R}^{j}}{x_{k}-x_{R}^{j}}\right)\\
\begin{array}{c}
\sqrt{(x_{k}-x_{T})+(y_{k}-y_{T})}+\\
\sqrt{(x_{k}-x_{R}^{j})+(y_{k}-y_{R}^{j})}
\end{array}
\end{array}\right]+\boldsymbol{\boldsymbol{\varepsilon}}_{k}^{j}
\end{array}\label{eq-35}
\end{equation}

\noindent where $t=[x_{T},y_{T}]^{\textrm{T}}$ is the location of
the transmitter and $\boldsymbol{\boldsymbol{\varepsilon}}_{k}^{j}\sim\mathcal{N}(0;0,R_{k}^{j})$
with $R_{k}^{j}=\textrm{diag}([\sigma_{\varphi}^{2},\sigma_{\rho}^{2}])$.
The scales of measurement noise are given by: $\sigma_{\varphi}=\varphi_{0}+\eta_{\varphi}\left\Vert r^{j}-p_{k}\right\Vert $;
$\sigma_{\rho}=\rho_{0}+\eta_{\rho}\left\Vert r^{j}-p_{k}\right\Vert ^{2}$,
in which $\left\Vert r^{j}-p_{k}\right\Vert ^{2}=(x_{R}^{j}-x_{k})^{2}+(y_{R}^{j}-y_{k})^{2}$,
$\rho_{0}=1\textrm{m}$, $\eta_{\rho}=5\times10^{-5}\textrm{ m}^{-1}$,
$\varphi_{0}=\pi/180\textrm{ rad}$, and $\eta_{\varphi}=1\times10^{-5}\textrm{ m}^{-1}$.
The geometry of a single transmitter-target-receiver pair is illustrated
in Fig. \ref{fig:The-geometry-of}, where locations of the transmitter,
receiver and target are denoted as $(x_{T},y_{T})$, $(x_{R},y_{R})$
and $(x,y)$, respectively. Clutter measurements for each receiver
are distributed uniformly over the region $[-\pi,\pi]\textrm{ rad}\times[0,15000]\textrm{ m}$.
The number of clutter measurements per scan is assumed to be Poisson
distributed with the clutter intensity $\lambda_{c}=2\times10^{-5}\textrm{ (\textrm{radm}})^{-1}$.
Without loss of generality, the tracking performance of the proposed
RFS-based approach is evaluated using the OSPA error distance \cite{40},
in which the order parameter $p$ determines the sensitivity to outliers
and the cut-off parameter $c$ determines the relative weighting of
the penalties assigned to cardinality and localization errors. For
each scenario, the average error performances are obtained over 50
MC trials. All experiments are tested in Matlab R2010a and implemented
on a computer with a 3.40 GHz processor. 

\subsection{Scenario 1}

\begin{figure}[htbp]
\noindent \begin{centering}
\includegraphics{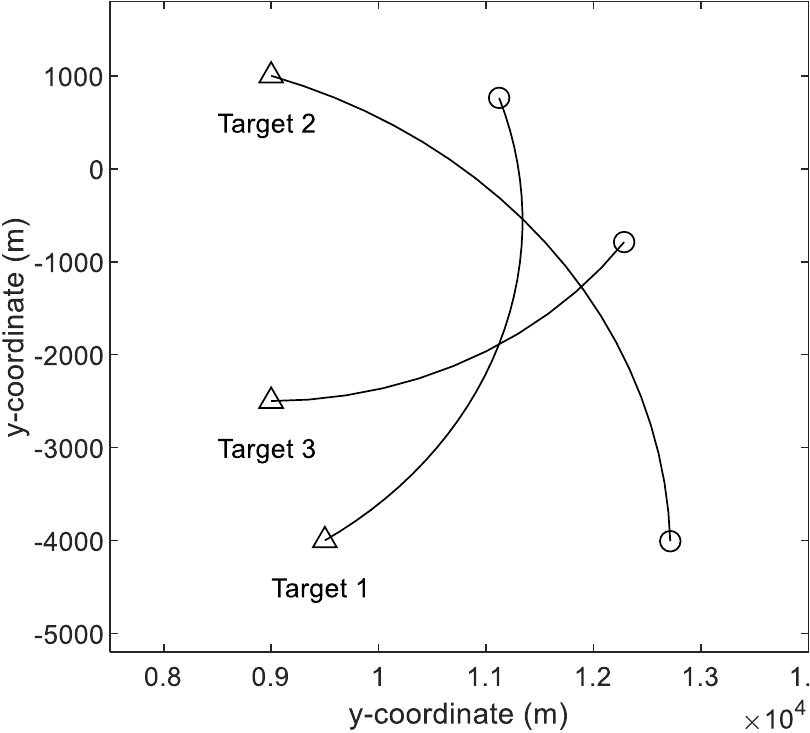}
\par\end{centering}

\caption{Target trajectories of Scenario 1 in the $x/y$ plane. Start/Stop
positions for each track are shown with $\circ/\vartriangle$. \label{fig:Target-trajectories-of}}
\end{figure}

In this scenario, a nearly constant turn model is considered. The
target state is denoted as $x_{k}=[\widetilde{x}_{k}^{\textrm{T}},\omega_{k}]^{\textrm{T}}$
where $\widetilde{x}_{k}=[x_{k},\dot{x}_{k},y_{k},\dot{y}_{k}]^{\textrm{T}}$
comprises the target position and velocity and $\omega_{k}$ is the
turn rate. The state transition model is 

\[
\widetilde{x}_{k}=F(\omega_{k-1})\widetilde{x}_{k-1}+G\omega_{k-1}
\]

\noindent where 

\[
F(\omega_{k-1})=\left[\begin{array}{cccc}
1 & \frac{\sin\omega T_{s}}{\omega} & 0 & -\frac{1-\cos\omega T_{s}}{\omega}\\
0 & \cos\omega T_{s} & 0 & -\sin\omega T_{s}\\
0 & \frac{1-\cos\omega T_{s}}{\omega} & 1 & \frac{\sin\omega T_{s}}{\omega}\\
0 & \sin\omega T_{s} & 0 & \cos\omega T_{s}
\end{array}\right]
\]

\[
G=\left[\begin{array}{cc}
\frac{T_{s}^{2}}{2} & 0\\
T_{s} & 0\\
0 & \frac{T_{s}^{2}}{2}\\
0 & T_{s}
\end{array}\right]
\]

\noindent and $\omega_{k-1}\sim\mathcal{N}(0;0,Q)$ is a 2D independent
and identically distributed Gaussian process noise vector with covariance
$Q=\sigma_{\omega}^{2}I_{2}$, where $\sigma_{\omega}=0.01\textrm{ m/}\textrm{\ensuremath{s^{2}}}$
is the standard deviation of the target acceleration. The sampling
interval is fixed to $T_{s}=10$ s. A total of three targets appear
in the surveillance area and the true trajectories of targets are
shown in Fig. \ref{fig:Target-trajectories-of}, in which Target 1
is born at $k=1$, Target 2 is born at $k=1$0, and Target 3 is born
at $k=20$. 

\begin{figure}[htbp]
\noindent \begin{centering}
\includegraphics{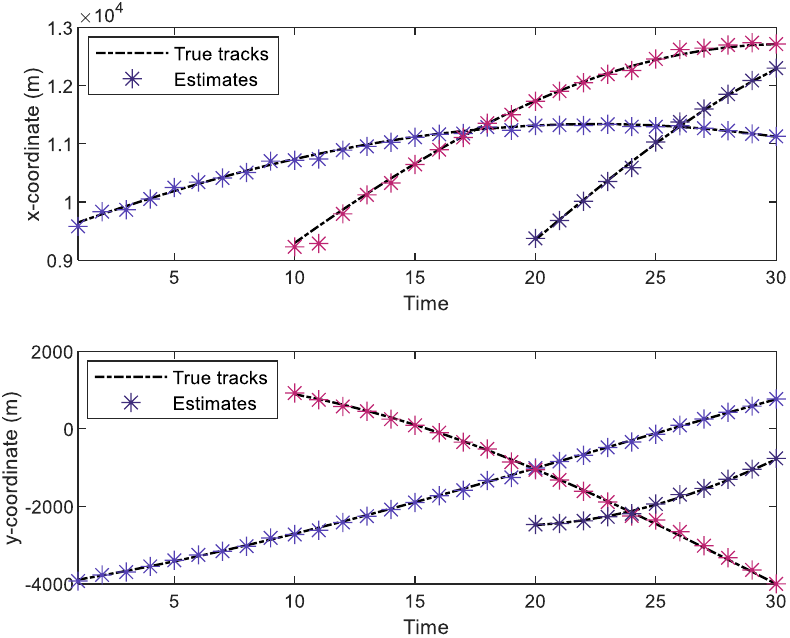}
\par\end{centering}

\caption{True and estimated tracks versus time in Scenario 1. \label{fig:True-and-estimated}}
\end{figure}

The birth process is a labeled multi-Bernoulli RFS with parameters
$f_{B}(x)=\{w_{B},p_{B}^{(i)}\}_{i=1}^{3}$, where the common existence
probabilities $w_{B}=0.02$, and $p_{B}^{(i)}(x)=\mathcal{N}(x;m_{B}^{(i)},P_{B})$
with $m_{B}^{(1)}=[9500,0,-4000,0,0]^{\textrm{T}}$, $m_{B}^{(2)}=[9000,0,1000,0,0]^{\textrm{T}}$,
$m_{B}^{(3)}=[9000,0,-2500,0,0]^{\textrm{T}}$, and $P_{B}=\textrm{diag}\text{(}[100,10,100,10,\pi/180)])^{2}$.
The units are meters for $x$ and $y$ and meters per second for $\dot{x}$
and $\dot{y}$. At each time step, $L=3000$ particles per birth track
is imposed. The number of components calculated and stored in each
forward propagation is set to be 1000. The survival probability is
fixed to $p_{s}=0.99$. The ESS-DFS-GLMB filter output for a single
MC run for the case where the number of sensors selected at each time
step $k$ is fixed to $P=|A_{k}|=2$, is given in Fig. \ref{fig:True-and-estimated},
showing the true and estimated tracks in $x$ and $y$ coordinates
versus time. The plots indicate that the ESS-DFS-GLMB filter is able
to identify all target births, as well as successfully accommodating
nonlinearities. 

To evaluate the performance of the ESS-DFS-GLMB filter, we compare
it with the heuristic random selection method, the exhaustive search
scheme, and the ESS-GLMB filter. In the random selection method, $P$
sensors are selected randomly at each time step and it is assumed
that each sensor has an equal probability to be chosen. In the exhaustive
search scheme, the Cauchy-Schwarz divergence is computed for all possible
combinations of $P$ sensors in the network and is used to solve the
optimization problem. In order to have an optimal solution, the order
of sensors is taken into account in the exhaustive search scheme.
To validate the performance of the dual-stage multi-sensor fusion
in the ESS-DFS-GLMB filter, the ESS-GLMB filter is also used as a
benchmark in which sensors are selected using the proposed efficient
selection method but measurements from selected sensors are updated
in a random order. To make the comparisons more meaningful, the PIMS
strategy is taken by all of the above methods.

\begin{figure}[htbp]
\noindent \begin{centering}
\includegraphics{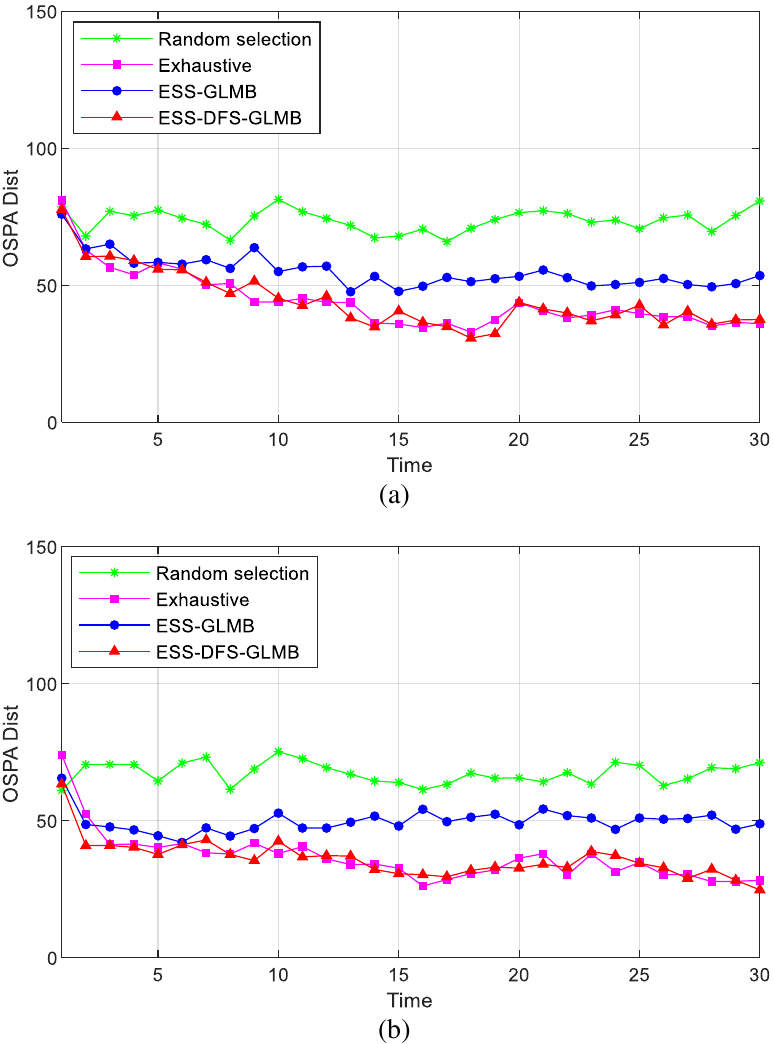}
\par\end{centering}

\caption{Average OSPA errors in Scenario 1: (a) $P=2$; (b) $P=3$. \label{fig:Average-OSPA-errors}}
\end{figure}

The MC averages of the OSPA distance (for $p=1$, $c=200$ m) for
two and three sensors selected at each time step are given in Fig.
\ref{fig:Average-OSPA-errors}(a) and (b), respectively. It can be
observed that the ESS-DFS-GLMB filter performs better than the random
selection method and the ESS-GLMB filter. In passive sensor networks,
the probability of detection of the passive sensor decreases quickly
the distance from the receiver increases. Therefore, the detection
abilities of sensors vary considerably with moving of targets. In
this case, the proposed dual-step multi-sensor fusion is a good choice
to improve the multi-sensor fusion results. Fig. \ref{fig:Average-OSPA-errors}
also demonstrates that in terms of multi-target tracking errors, the
ESS-DFS-GLMB filter is even comparable with the state of art.

To compare the computational efficiencies of the ESS-DFS-GLMB filter
and the exhaustive search scheme, the average computing time (in sec)
for each method to execute a complete MC simulation against different
number $P$ of selected sensors is shown in Table 1. As seen from
Table 1, there is an overall increase of the computing time with the
increase of $P$. When $P=1$, the computing time of the exhaustive
search scheme and the ESS-DFS-GLMB method is comparable with each
other. As mentioned before, when $P=1$, these two methods degenerate
into the same sensor management solution. The ESS-DFS-GLMB filter
runs about 13.45 times faster than the exhaustive search method when
$P=2$, and 137.33 times faster when $P=3$. These results are good
agreement with the theoretical analysis of the computational complexity.

\begin{table}[htbp]
\caption{Computing time comparison for Scenario 1 (s)}

\centering{}%
\begin{tabular}{|c|c|c|}
\hline 
Number of selected sensors & Exhaustive & ESS-DFS-GLMB\tabularnewline
\hline 
\hline 
$P=1$ & 181.45 & 176.72\tabularnewline
\hline 
$P=2$ & 3908.18 & 290.47\tabularnewline
\hline 
$P=3$ & 52161.40 & 379.83\tabularnewline
\hline 
\end{tabular}
\end{table}

\subsection{Scenario 2}

We now consider a more complicated scenario and assume a total number
of six targets in the surveillance area. The target state variable
is a vector of target position and velocity and is denoted as $x_{k}=[x_{k},y_{k},\dot{x}_{k},\dot{y}_{k}]$.
The single-target transition model is linear Gaussian specified by 

\[
x_{k}=F_{k}x_{k-1}+u_{k}
\]

\noindent with the transition matrix 

\[
F{}_{k}=\left[\begin{array}{cccc}
1 & 0 & T_{s} & 0\\
0 & 1 & 0 & T_{s}\\
0 & 0 & 1 & 0\\
0 & 0 & 0 & 1
\end{array}\right].
\]

\begin{figure}[htbp]
\noindent \begin{centering}
\includegraphics{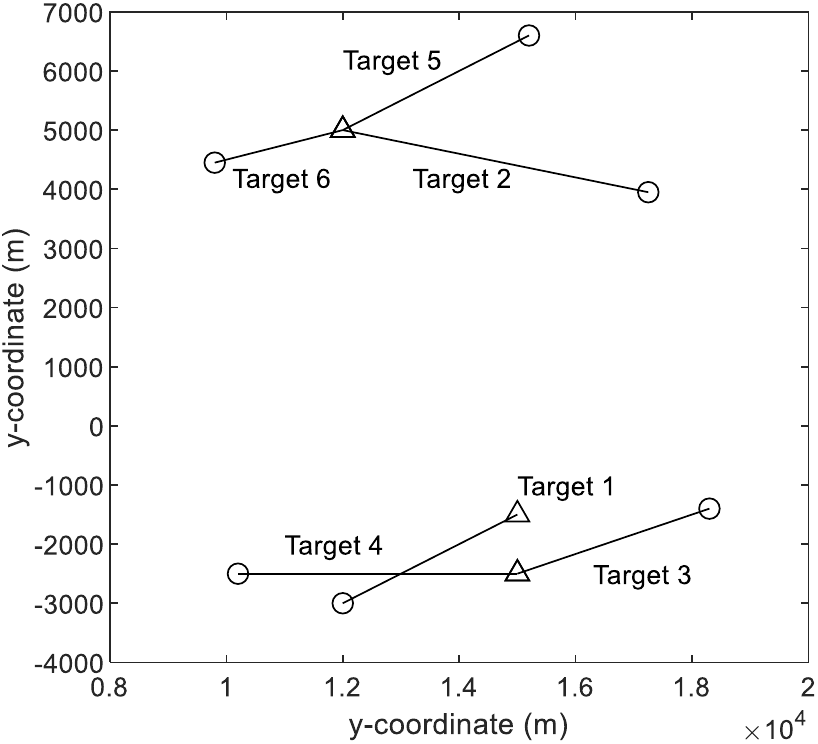}
\par\end{centering}

\caption{Target trajectories of Scenario 2 in the $x/y$ plane. Start/Stop
positions for each track are shown with $\circ/\vartriangle$. \label{fig:Target-trajectories-of-1}}
\end{figure}

\noindent The process noise is zero-mean Gaussian distributed with
covariance 

\[
Q=\sigma_{u}^{2}\left[\begin{array}{cccc}
\frac{T_{s}^{3}}{3} & 0 & \frac{T_{s}^{2}}{2} & 0\\
0 & \frac{T_{s}^{3}}{3} & 0 & \frac{T_{s}^{2}}{2}\\
\frac{T_{s}^{2}}{2} & 0 & T_{s} & 0\\
0 & \frac{T_{s}^{2}}{2} & 0 & T_{s}
\end{array}\right]
\]

\noindent where $\sigma_{u}=0.01\textrm{\textrm{ m/}\ensuremath{s^{2}}}$
is the standard deviation of the process noise. The true trajectories
are shown in Fig. \ref{fig:Target-trajectories-of-1}, in which Target
1 is born at $k=1$, Targets 2 is born at $k=10$, Targets 3 and 6
are born at $k=20$, and Targets 4 and 5 are born at $k=15$. 

\begin{figure}[htbp]
\noindent \begin{centering}
\includegraphics{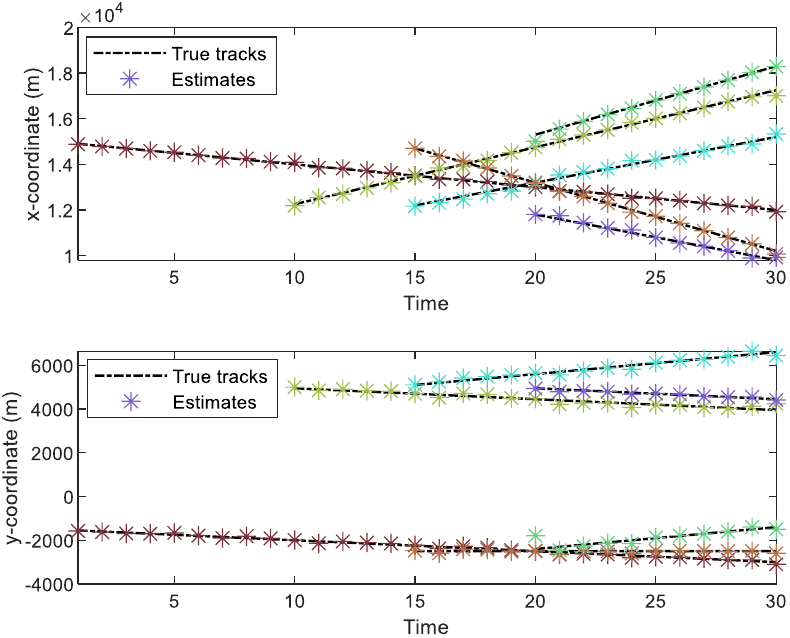}
\par\end{centering}

\caption{True and estimated tracks versus time in Scenario 2. \label{fig:True-and-estimated-1} }
\end{figure}

The birth process is a labeled multi-Bernoulli RFS with parameters
$f_{B}(x)=\{w_{B},p_{B}^{(i)}\}_{i=1}^{3}$, where the common existence
probabilities $w_{B}=0.02$, and $p_{B}^{(i)}(x)=\mathcal{N}(x;m_{B}^{(i)},P_{B})$
with $m_{B}^{(1)}=[15000,0,-1500,0]^{\textrm{T}}$, $m_{B}^{(2)}=[12000,0,5000,0]^{\textrm{T}}$,
$m_{B}^{(3)}=[15000,0,-2500,0]^{\textrm{T}}$, and $P_{B}=\textrm{diag}\text{(}[100,10,100,10]^{\textrm{T}})^{2}$.
All other parameters are the same as those in the previous scenario.
The ESS-DFS-GLMB filter output for a single MC run for the case where
the number of sensors selected at each time step $k$ is fixed to
$P=|A_{k}|=2$, is given in Fig. \ref{fig:True-and-estimated-1},
showing the true and estimated tracks in $x$ and $y$ coordinates
versus time. The results suggest that the trajectory estimates are
very close to the true target trajectories.

\begin{figure}[htbp]
\noindent \begin{centering}
\includegraphics{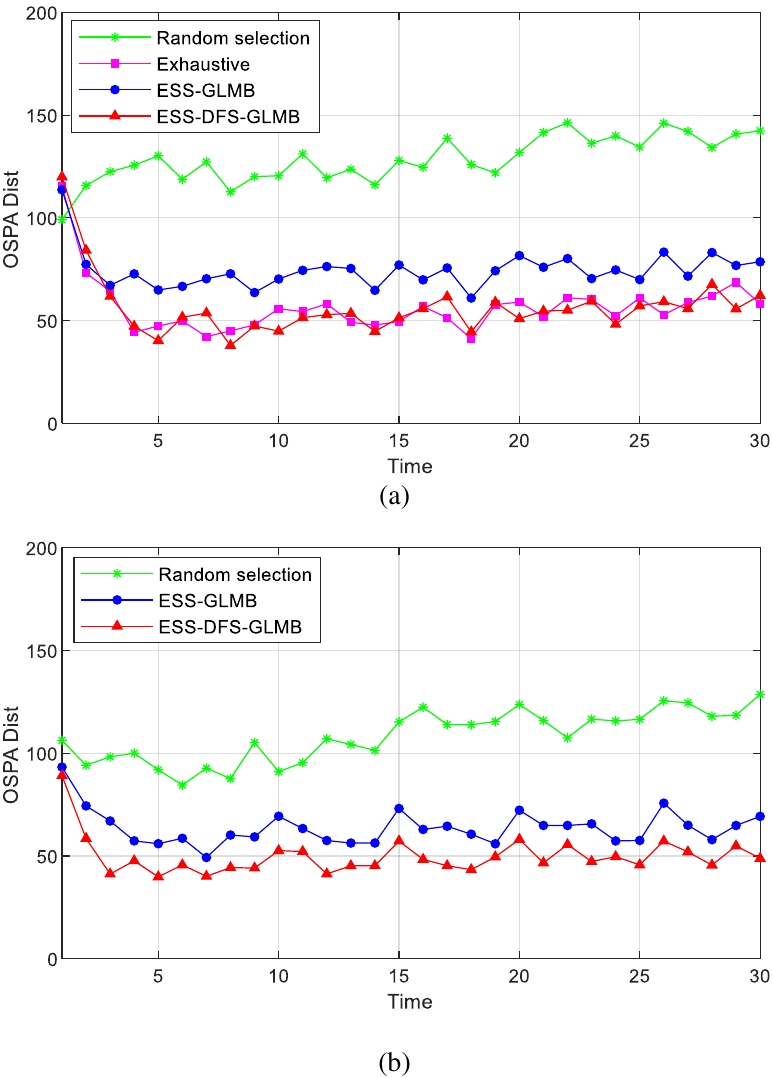}
\par\end{centering}

\caption{Average OSPA errors in scenario 2 (a) $P=2$; (b) $P=3$. \label{fig:Average-OSPA-errors-1}}
\end{figure}

The MC averages of the OSPA distance (for $p=1$, $c=200$ m) for
two and three sensors selected at each time step are given in Fig.
\ref{fig:Average-OSPA-errors-1}(a) and (b), respectively. When P=3,
the computing burden of the exhaustive search scheme is overload and
hence it is not considered here. It can be observed from Fig. \ref{fig:Average-OSPA-errors-1}
that the ESS-DFS-GLMB filter performs better than the random selection
method and the ESS-GLMB filter. When $P=2$, the tracking accuracy
of the ESS-DFS-GLMB filter and the optimal exhaustive search scheme
is comparable with each other.

To compare the computational efficiencies of the ESS-DFS-GLMB filter
and the exhaustive search scheme, the average computing time (in sec)
for each method to execute a complete MC simulation against different
number $P$ of selected sensors is shown in Table 2. The computing
time of the exhaustive method and the ESS-DFS-GLMB filter is comparable
with each other when $P=1$. The ESS-DFS-GLMB filter runs about 16.89
times faster than the exhaustive search method when $P=2$.

\begin{table}[htbp]
\caption{Computing time comparison for Scenario 2 (s)}

\centering{}%
\begin{tabular}{|c|c|c|}
\hline 
Number of selected sensors & Exhaustive & ESS-DFS-GLMB\tabularnewline
\hline 
\hline 
$P=1$ & 301.49 & 289.71\tabularnewline
\hline 
$P=2$ & 9349.23 & 553.47 \tabularnewline
\hline 
$P=3$ & Overload & 792.10\tabularnewline
\hline 
\end{tabular}
\end{table}

\section{Conclusions}

An efficient sensor management approach named ESS-DFS-GLMB has been
proposed in this paper for multi-target tracking in passive sensor
networks. Since the number of sensors in passive sensor networks is
generally large, it would not be feasible to directly use the information
on the entire sensor collection and hence the need for sensor selection.
However, the sensor selection is a global combinatorial optimization
problem that is extremely challenging when the network is large. To
solve this problem, an efficient information-theoretic sensor selection
strategy has been developed using the GLMB filter. At each time step,
a fixed number of sensors are selected sequentially from the candidates
based on the Cauchy-Schwarz divergence. For the selected sensors,
the iterated-corrector multi-sensor GLMB filter is applied sequentially
in an order determined by their Cauchy-Schwarz divergences. This so-called
proposed dual-stage sequential update scheme well improves the fusion
accuracy without introducing any additional computation.

Numerical studies are presented for two challenging scenarios where
multiple moving targets are to be tacked in a passive sensor network
using non-linear measurements. The results demonstrate the superior
tracking accuracy of the ESS-DFS-GLMB approach in terms of OSPA errors.
What\textquoteright s more, the computing time comparison results
show that it runs substantially faster than the exhaustive search-based
technique. These results are good agreement with the theoretical analysis
of the computational complexity. Future work will consider the learning
of clutter statistics and integration of data over multiple scans
\cite{12} to improve tracking performance.

\bibliographystyle{IEEEtran}
\bibliography{Efficient_Sensor_Management_for_Multitarget_Tracking_in_Passive_Sensor_Networks_via_Cauchy-Schwarz_Divergence}

\begin{thebibliography}{10}
\providecommand{\url}[1]{#1}
\csname url@samestyle\endcsname
\providecommand{\newblock}{\relax}
\providecommand{\bibinfo}[2]{#2}
\providecommand{\BIBentrySTDinterwordspacing}{\spaceskip=0pt\relax}
\providecommand{\BIBentryALTinterwordstretchfactor}{4}
\providecommand{\BIBentryALTinterwordspacing}{\spaceskip=\fontdimen2\font plus
\BIBentryALTinterwordstretchfactor\fontdimen3\font minus
  \fontdimen4\font\relax}
\providecommand{\BIBforeignlanguage}[2]{{%
\expandafter\ifx\csname l@#1\endcsname\relax
\typeout{** WARNING: IEEEtran.bst: No hyphenation pattern has been}%
\typeout{** loaded for the language `#1'. Using the pattern for}%
\typeout{** the default language instead.}%
\else
\language=\csname l@#1\endcsname
\fi
#2}}
\providecommand{\BIBdecl}{\relax}
\BIBdecl

\bibitem{1}
S.~Joshi and S.~Boyd, ``Sensor selection via convex optimization,'' \emph{IEEE
  Trans. Signal Proces.}, vol.~57, no.~2, pp. 451--462, 2009.

\bibitem{2}
J.~Yick, B.~Mukherjee, and D.~Ghosal, ``Wireless sensor network survey,''
  \emph{Comput. Netw.}, vol.~52, no.~12, pp. 2292--2330, 2008.

\bibitem{3}
R.~Mahler, \emph{Statistical Multisource-Multitarget Information Fusion}.\hskip
  1em plus 0.5em minus 0.4em\relax Norwood, MA: Artech House, 2007.

\bibitem{4}
{R. Mahler}, ``Multi-target {Bayes} filtering via first-order multi-target
  moments,'' \emph{IEEE Trans. Aerosp. Electron. Syst.}, vol.~39, no.~4, pp.
  1152--1178, 2003.

\bibitem{5}
R.~{Mahler}, ``{PHD} filters of higher order in target number,'' \emph{IEEE
  Trans. Aerosp. Electron. Syst.}, vol.~43, no.~4, pp. 1523--1543, 2007.

\bibitem{6}
B.-T. Vo, B.-N. Vo, and A.~Cantoni, ``The cardinality balanced multi-target
  {multi-Bernoulli} filter and its implementations,'' \emph{IEEE Trans. Signal
  Proces.}, vol.~57, no.~2, pp. 409--423, 2009.

\bibitem{7}
B.-T. Vo and B.-N. Vo, ``Labeled random finite sets and multi-target conjugate
  priors,'' \emph{IEEE Trans. Signal Proces.}, vol.~61, no.~13, pp. 3460--3475,
  2013.

\bibitem{8}
B.-N. Vo, B.-T. Vo, and D.~Phung, ``Labeled random finite sets and the {Bayes}
  multi-target tracking filter,'' \emph{IEEE Trans. Signal Proces.}, vol.~62,
  no.~24, pp. 6554--6567, 2014.

\bibitem{9}
B.-N. Vo, B.-T. Vo, and H.~G. Hoang, ``An efficient implementation of the
  generalized labeled {multi-Bernoulli} filter,'' \emph{IEEE Trans. Signal
  Proces.}, vol.~65, no.~8, pp. 1975--1987, 2017.

\bibitem{10}
M.~Beard, B.-T. Vo, and B.-N. Vo, ``A solution for large-scale multi-object
  tracking,'' \emph{IEEE Trans. Signal Proces.}, vol.~68, pp. 2754--2769, 2020.

\bibitem{11}
B.-N. Vo, B.-T. Vo, and M.~Beard, ``Multi-sensor multi-object tracking with the
  generalized labeled {multi-Bernoulli} filter,'' \emph{IEEE Trans. Signal
  Proces.}, vol.~67, no.~23, pp. 5952--5967, 2019.

\bibitem{12}
B.-N. Vo and B.-T. Vo, ``Multi-scan generalized labeled {multi-Bernoulli}
  models for multi-object state estimation,'' \emph{IEEE Trans. Signal
  Proces.}, vol.~67, no.~19, pp. 4998--4963, 2019.

\bibitem{13}
H.~G. Hoang and B.-T. Vo, ``Sensor management for multi-target tracking via
  {multi-Bernoulli} filtering,'' \emph{Automatica}, vol.~50, no.~4, pp.
  1135--1142, 2014.

\bibitem{14}
A.~K. Gostar, R.~Hoseinnezhad, and A.~Bab-Hadiashar, ``{Multi-Bernoulli} sensor
  control for multi-target tracking,'' in \emph{Proc. Int. Con. Intelligent.
  Sensors, Sensor Networks and Information Processing (ISSNIP 2013)},
  Melbourne, Australia, 2013, pp. 312--317.

\bibitem{15}
A.~K. Gostar, R.~Hoseinnezhad, and A.~{Bab-Hadiashar}, ``{Multi-Bernoulli}
  sensor control via minimization of expected estimation errors,'' \emph{IEEE
  Trans. Aerosp. Electron. Syst.}, vol.~51, no.~3, pp. 1762--1773, 2015.

\bibitem{16}
{R. Mahler}, ``Multitarget sensor management of dispersed mobile sensors,'' in
  \emph{Theory and Algorithms for Cooperative Systems}, D.~Grundel, R.~Murphey,
  and P.~Pardalos, Eds.\hskip 1em plus 0.5em minus 0.4em\relax Singapore: World
  Scientific, 2004, pp. 239--310.

\bibitem{17}
R.~Mahler, ``Sensor management with non-ideal sensor dynamics,'' in \emph{Proc.
  7th Int. Conf. Information Fusion}, Stockholm, Sweden, 2004, pp. 1--8.

\bibitem{18}
{R. Mahler}, ``Unified sensor management using {CPHD} filters,'' in \emph{Proc.
  10th Int. Conf. Information Fusion}, Quebec City, Canada, 2007, pp. 1--7.

\bibitem{19}
A.~K. Gostar, R.~Hoseinnezhad, and A.~{Bab-Hadiashar}, ``Robust multi-bernoulli
  sensor selection for multi-target tracking in sensor networks,'' \emph{IEEE
  Signal Process. Lett.}, vol.~20, no.~12, pp. 1167--1170, 2013.

\bibitem{20}
A.~K. Gostar, R.~Hoseinnezhad, and A.~Bab{-Hadiashar}, ``Multi-{B}ernoulli
  sensor-selection for multi-target tracking with unknown clutter and detection
  profiles,'' \emph{Signal Process.}, vol. 119, pp. 28--42, 2016.

\bibitem{21}
A.~K. Gostar, R.~Hoseinnezhad, and A.~Bab-Hadiashar, ``Sensor control for
  multi-object tracking using labeled {multi-Bernoulli} filter,'' in
  \emph{Proc. Int. Conf. Information Fusion}, Salamanca, 2014, pp. 1--8.

\bibitem{22}
Y.~Zhu, J.~Wang, and S.~Liang, ``Multi-objective optimization based
  {multi-Bernoulli} sensor selection for multi-target tracking,''
  \emph{Sensors}, vol.~19, no.~4, pp. 1--18, 2019.

\bibitem{23}
{R. Mahler}, ``Global posterior densities for sensor management,'' in
  \emph{Proc. International Society for Optics and Photonics}, Orlando, USA,
  1998, pp. 252--263.

\bibitem{24}
B.~Ristic and B.-N. Vo., ``Sensor control for multi-object state-space
  estimation using random finite sets,'' \emph{Automatica}, vol.~46, no.~11,
  pp. 1812--1818, 2010.

\bibitem{25}
H.~G. Hoang, B.-N. Vo, B.-T. Vo, and R.~Mahler, ``The {Cauchy-Schwarz}
  divergence for {Poisson} point processes,'' \emph{IEEE Trans. Information
  Theory}, vol.~61, no.~8, pp. 4475--4485, 2015.

\bibitem{26}
M.~Beard, B.-T. Vo, B.-N. Vo, and S.~Arulampalam, ``Void probabilities and
  {Cauchy-Schwarz} divergence for generalized labeled {multi-Bernoulli}
  models,'' \emph{IEEE Trans. Signal Proces.}, vol.~65, no.~19, pp. 5047--5061,
  2017.

\bibitem{27}
H.~V. Nguyen, H.~Rezatofighi, B.-N. Vo, and D.~Ranasinghe, ``Online {UAV} path
  planning for joint detection and tracking of multiple radio-tagged objects,''
  \emph{IEEE Trans. Signal Proces.}, vol.~67, no.~20, pp. 5365--5379, 2019.

\bibitem{28}
H.~V. {Nguyen}, H.~Rezatofighi, B.-N. Vo, and D.~Ranasinghe, ``Multi-objective
  multi-agent planning for jointly discovering and tracking mobile object,'' in
  \emph{Proc. AAAI Conference on Artificial Intelligence}, New York, USA, 2020,
  pp. 7227--7235.

\bibitem{29}
L.~Ma, K.~Xue, and P.~Wang, ``Multitarget tracking with spatial nonmaximum
  suppressed sensor selection,'' \emph{Mathematical Problems in Engineering},
  vol.~4, pp. 1--10, 2015.

\bibitem{30}
L.~{Ma}, K.~Xue, and P.~Wang, ``Distributed multiagent control approach for
  multitarget tracking,'' \emph{Mathematical Problems in Engineering}, vol.~1,
  pp. 1--10, 2015.

\bibitem{31}
P.~Wang, L.~Ma, and K.~Xue, ``Multitarget tracking in sensor networks via
  efficient information-theoretic sensor selection,'' \emph{International
  Journal of Advanced Robotic Systems}, vol.~14, no.~5, pp. 1--9, 2017.

\bibitem{32}
H.~D. Griffiths and C.~J. Baker, ``Passive coherent location radar systems.
  {Part} 1: {Performance} prediction,'' \emph{IET Radar Sonar and Navigation},
  vol.~3, no. 152, pp. 153--159, 2005.

\bibitem{33}
K.~Chetty, K.~Woodbridge, H.~Guo, and G.~Smith, ``Passive bistatic {WiMAX}
  radar for marine surveillance,'' in \emph{IEEE Radar Conference}, Washington,
  DC, USA, 2010, pp. 188--193.

\bibitem{34}
M.~Daun and U.~Nickel, ``Tracking in multistatic passive radar systems using
  {DAB/DVB-T} illumination,'' \emph{Signal Process.}, vol.~92, no.~6, pp.
  1365--1386, 2012.

\bibitem{35}
D.~K.~P. Tan, H.~Sun, Y.~Lu, M.~Lesturgie, and H.~L. Chan, ``Passive radar
  using global system for mobile communication signal: {Theory}, implementation
  and measurements,'' \emph{IET Radar Sonar and Navigation}, vol.~3, no. 152,
  pp. 116--123, 2005.

\bibitem{36}
T.~S. Yoo and S.~Lafortune, ``{NP}-completeness of sensor selection problems
  arising in partially observed discrete-event systems,'' \emph{IEEE Trans.
  Autom. Control}, vol.~47, no.~9, p. 1495–1499, 2002.

\bibitem{37}
F.~Bian, D.~Kempe, and R.~Govindan, ``Utility based sensor selection,'' in
  \emph{Proc. 5th Int. Conf. Information Processing in Sensor Networks}, New
  York, USA, 2006, pp. 11--18.

\bibitem{38}
B.~Ristic and A.~Farina, ``Target tracking via multi-static {Doppler} shifts,''
  \emph{IET Radar Sonar and Navigation}, vol.~7, no.~5, pp. 508--516, 2013.

\bibitem{39}
B.~R. Mahafza, \emph{Radar systems analysis and design using {MATLAB}, 3rd
  ed.}\hskip 1em plus 0.5em minus 0.4em\relax Boca Raton, Florida, USA: Chapman
  and Hall/CRC Press., 2013.

\bibitem{40}
D.~Schuhmacher, B.-T. Vo, and B.-N. Vo, ``A consistent metric for performance
  evaluation of multi-object filters,'' \emph{IEEE Trans. Signal Process.},
  vol.~56, no.~8, pp. 3447--3457, 2008.

\end{thebibliography}

\end{document}